\documentclass[preprint,12pt,3p]{elsarticle}
\usepackage[normalem]{ulem}
\usepackage{caption}
\usepackage{subcaption}
\usepackage{dcolumn}
\usepackage{bm}
\usepackage{color}
\usepackage{hyperref}
\usepackage{gensymb}
\usepackage{amsmath}
\usepackage{empheq}
\usepackage{booktabs}
\usepackage{multirow}
\usepackage{graphicx}
\usepackage{tabu}
\usepackage{mhchem}
\usepackage{soul,xcolor}
\setstcolor{red}

\journal{Acta Materialia}

\begin{document}

\begin{frontmatter}

\title{The bi-layered precipitate phase $\zeta$ in the Al-Ag alloy system}

\author[MatEng]{Zezhong Zhang}
\author[MatEng,MCEM]{Laure Bourgeois\corref{cor1}}
\ead{laure.bourgeois@monash.edu}
\author[Erich]{Julian M. Rosalie}
\author[MatEng]{Nikhil V. Medhekar\corref{cor2}}
\ead{nikhil.Medhekar@monash.edu}

\address[MatEng]{Department of Materials Science and Engineering, Monash University, Victoria 3800, Australia}
\address[MCEM]{Monash Centre for Electron Microscopy, Monash University, Victoria 3800, Australia}
\address[Erich]{Erich Schmid Institute of Materials Science, Austria}
\cortext[cor1]{Corresponding author}
\cortext[cor2]{Corresponding author}

\begin{abstract}
The Al-Ag system is thought to be a well-understood model system used to study diffusional phase transformations in alloys. Here we report the existence of a new precipitate phase, $\zeta$, in this classical system using scanning transmission electron microscopy (STEM). The $\zeta$ phase has a modulated structure composed of alternative bilayers enriched in Al or Ag. Our {\em in situ} annealing experiments reveal that the $\zeta$ phase is an intermediate precipitate phase between GP zones and $\gamma^\prime$. First-principles calculations show that $\zeta$ is a local energy minimum state formed during Ag clustering in Al. The layered structure of $\zeta$ is analogous to the well-known Ag segregation at the precipitate-matrix interfaces when Ag is microalloyed in various aluminium alloys.
\end{abstract}


\end{frontmatter}



\section{Introduction}
Al-Ag alloys have been studied extensively since last century and now serve as a textbook alloy system \cite{Guinier1942,porter2011,Aaronson2010,Askeland2010}. It is a model system to study solid-solid phase transformations for several reasons. First, one of the transformations associated with the decomposition of the supersaturated solid solution involves a structural change from face-centred cubic (FCC) to hexagonal close-packed (HCP) that is straightforward to understand \cite{porter2011,Aaronson2010}. Second, the atomic size difference between Al and Ag is negligible, which gives a minimal volumetric strain associated with the solute clustering and phase transformations \cite{dubey1991}. In addition, the large difference between the atomic numbers ($Z_\mathrm{Al}=13$ and $Z_\mathrm{Ag}=47$) is particularly favourable for Z-contrast imaging in the transmission electron microscope \cite{moore2000,erni2003a,erni2003b}. Thus, Al-Ag alloys are often chosen to demonstrate advanced electron microscopy techniques, such as {\em in situ} annealing to observe the atomic mechanisms of precipitates growth in real time \cite{howe1990,howe1995} and electron tomography to reconstruct embedded precipitates with high spatial resolution \cite{inoke2006,Aert2011}.\par

The precipitation sequence in the Al-Ag alloy system is commonly recognised as \cite{porter2011,Aaronson2010}:
\begin{center}
$\alpha^\prime$ $\rightarrow$ GP zones ($\epsilon$ or $\eta$) $\rightarrow$ $\gamma^\prime$/$\gamma$,
\end{center}
where $\alpha^\prime$ represents the supersaturated solid solution. Guinier-Preston (GP) zones are the early-stage solute enriched regions. Unlike most aluminium alloys, GP zones in the Al-Ag system form immediately after quenching \cite{baur1962}. Given an asymmetric miscibility gap \cite{baur1962}, two types of GP zones are proposed, with their compositions depending on ageing temperature: $\epsilon$ forms at high temperature ($>$ 170 \degree C) with relatively low Ag concentration (below 44 at.\%) while $\eta$ forms at low temperature ($<$ 170 \degree C) with relatively high Ag concentration (44 at.\%-60 at.\%) \cite{baur1962,Gragg1971,al1993}. GP zone $\eta$ is thought to be uniform in composition while GP zone $\epsilon$ is believed to have a core-shell structure and there has been much debate as to whether Ag enriches the core or the shell \cite{dubey1991,guinier1996}. The $\gamma^\prime$ phase is a metastable precipitate phase formed before the equilibrium $\gamma$ phase. Both phases display the same composition (\ce{Ag2Al}) and the same atomic structure (HCP with space group $P6_3/mmc$) but with slightly different lattice parameters \cite{nicholson1961,neumann1966}. The nucleation and growth of $\gamma^\prime$ require Shockley partial dislocations to accommodate the shear associated with the FCC to HCP transformation \cite{howe1987}. However, Al has a particularly high stacking fault energy \cite{Ogata_Ideal_2002}, which means HCP precipitates are generally hard to nucleate. The difficult nucleation inevitably gives a low precipitate number density of $\gamma^\prime$/$\gamma$. As a consequence, Al-Ag alloys have a poor mechanical performance \cite{rosalie2009}.

Though binary Al-Ag alloys only have limited structural applications, Ag is a popular microalloying element in aluminium alloys. A small amount (from 0.1 at.\% to 0.5 at.\%) of Ag was found to exert dramatic improvements on the mechanical properties, thermal stability and stress-corrosion cracking resistance of aluminium alloys \cite{polmear1960,auld-polmear1966}. This effect is widely seen in aluminium systems, particularly in the high-strength alloys for advanced aerospace and defence applications, including Al-Cu-Mg based alloys and Al-Zn-Mg based alloys \cite{Polmear2005,Sha2010}. The underlying mechanisms, however, seem different from case to case. For instance, Ag incorporates within existing precipitate phases and accelerates their nucleation kinetics in Al-Zn-Mg based alloys \cite{Ringer_Microstructural_2000} or Al-Mg-Si based alloys \cite{nakamura2010effect,Marioara_HAADF_2012,wenner2014atomic}. But in Al-Cu based alloys, Ag segregates at the precipitate-matrix interfaces, which changes precipitation behaviour and also modifies the type of precipitates that form \cite{Polmear_Role_1987}. Field ion microscopy \cite{hono1994pre,ringer1994precipitate} and atom probe tomography \cite{murayama1998three,Reich1998} have shown that Ag clusters with other solute elements (particularly Mg) at the start of ageing and segregates to $\{111\}_\mathrm{Al}$ planes, thus, initiating the $\Omega$ phase but suppressing the S phase in Al-Cu-Mg-Ag alloys. Positron annihilation lifetime spectroscopy (PALS) has also suggested that Ag binds with Mg, Cu and vacancies during ageing \cite{Somoza_Positron_2000}. Studies by scanning transmission electron microscopy (STEM) have confirmed that Ag segregation is via one or two layers at the precipitate-matrix interfaces for various precipitates with different alloying compositions, including $\Omega$ in Al-Cu-Mg-Ag \cite{Hutchinson2001,Kang2014}, $\theta'$ in Al-Cu-Ag \cite{Rosalie2012} and $T_1$ in Al-Cu-Li-Mg-Ag \cite{Kang2015}. Ag segregation at interfaces is believed to lower the interfacial energy \cite{Kang2014,Rosalie2012}. Ag atoms are also suspected to cluster on $\{111\}_\mathrm{Al}$ planes whereas Cu GP zones form on $\{001\}_\mathrm{Al}$ planes \cite{Baba_Influence_1969}. However, what drives Ag to cluster before formation of a precipitate is still a mystery. Moreover, how the early-stage clustering modifies the nucleation of precipitates is largely unknown. Accurate descriptions of solute clustering are essential to address one of the most intriguing questions in Al alloys: why does the minor addition of Ag play a crucial role in precipitation in a wide variety of Al alloy systems \cite{Polmear_Role_1987}?

With the evidence that Ag facilitates phase transformations for a broad range of aluminium alloys, we hypothesise that this is due to some intrinsic properties of Ag in aluminium. Therefore, we revisited the Al-Ag system and characterised the atomic structures of the different precipitate phases using scanning transmission electron microscopy (STEM). In particular, we examined the ordering of GP zones and $\gamma^\prime$ precipitates. Surprisingly, we found a new metastable precipitate phase consisting of a bi-layered structure, which we named $\zeta$. Based on our experiments and first-principles calculations, $\zeta$ is a metastable phase that forms before transforming into HCP phases. The layered structure of $\zeta$ on $\{111\}_\mathrm{Al}$ planes is analogous to Ag segregation at the precipitates-matrix interfaces. We find that Ag naturally prefers to decompose from the supersaturated solid solution and aggregate on $\{111\}_\mathrm{Al}$ planes.

\section{Experimental and Computational Methods}

\subsection{Sample Preparation}

The alloy composition used in this work was Al-1.68 at.\% Ag, as cast from high-purity aluminium (Cerac alloys, 99.99\% purity) and silver (AMAC alloys, 99.9+\%). The pure metals were melted in air at 700\degree C in a graphite crucible, stirred and poured into graphite-coated steel moulds. The compositions were measured by inductively coupled plasma atomic emission spectrometry, showing very low levels of impurities \cite{rosalie2014chemical}. The cast ingots were homogenised at 525\degree C for 7 days, then hot- and cold-rolled to 0.5 mm alloys sheets. The samples were in the form of disks 3 mm in diameter and 0.5 mm in thickness, punched from an alloy sheet after rolling. They were solutionised at 525\degree C for 30 min in a nitrate salt bath and quenched to room temperature. Different quenching media, including water and oil, were tested to manipulate the quenched-in vacancy concentration before ageing. Then the samples were aged at 200\degree C in an oil bath for a range of times (from 30 min to 7 days). The TEM specimens were made by mechanically grinding the disks and electro-polishing them in a 67\% methanol-33\% nitric acid mixture at -25\degree C and 13 V with a current average of 200 mA.\par

\subsection{Electron Microscopy}

The alloy microstructure and precipitate atomic structures were characterised by scanning transmission electron microscopy (STEM). In particular, high-angle annular dark-field (HAADF) STEM was performed to exploit the large difference in the atomic numbers between Ag and Al. Preliminary investigations of the microstructures were carried out on a JEOL JEM 2100F field-emission gun transmission electron microscope (FEGTEM) and a FEI Tecnai G$^\mathrm{2}$ F20 Super-Twin lens FEGTEM. The JEOL 2100F was operated at 200 kV and has a STEM probe size of 2 \AA. The semi-convergence angle used was 10 mrad and the HAADF detector had an inner collection semi-angle of 65 mrad and an outer collection semi-angle of 185 mrad. The {\em in situ} annealing experiments were conducted in the JEOL 2100F using a Gatan 652 double-tilt heating holder at various temperatures (100\degree C, 150\degree C and 200\degree C) for a short amount of time, ranging from 3 min to 70 min. In these experiments, samples were imaged either during {\em in situ} annealing or after cooling down to room temperature. The Tecnai F20 was operated at 200 kV and also has a STEM probe size of about 2 \AA. The semi-convergence angle used was 9.3 mrad and the HAADF detector had an inner collection semi-angle of 41 mrad and an outer collection semi-angle of 220 mrad. A tilt series was performed in the Tecnai F20 using a Fischione model 2020 single-tilt axis high-tilt sample holder. Higher resolution HAADF-STEM imaging was conducted in a dual-aberration-corrected FEI Titan$^3$ FEGTEM. The Titan$^3$ was operated at 300 kV and a convergence semi-angle of 15 mrad, which gave a probe size of about 1.2 \AA. HAADF imaging used an inner collection semi-angle of 55 mrad and an outer collection semi-angle of 200 mrad. BF imaging used an inner collection semi-angle of 13 mrad. The spherical aberration coefficient $C_s$ is about 1 $\mu$m in this Titan$^3$.\par

Compositional analysis was performed on the JEOL 2100F and the Tecnai F20 using energy-dispersive X-ray spectroscopy (EDS). The JEOL 2100F has a JEOL 50 mm$^\mathrm{2}$ Si(Li) detector with ultra-thin window. The Tecnai F20 has a Bruker XFlash 6120T 30 mm$^\mathrm{2}$ silicon drift windowless detector. The composition measured by EDS across the sample thickness is contributed by the Ag enriched precipitate and the Al matrix. The precipitates examined in this study have a roughly spherical geometry. Thus, the diameter of a precipitate in the electron transmission direction is approximated to be equal as its averaged diameter on the HAADF image. The thickness of the specimen in the vicinity of a precipitate was determined by comparing on-zone position-averaged convergent-beam electron diffraction (PACBED) of the surrounding matrix \cite{LeBeau_Position_2010} obtained experimentally and calculated PACBED patterns with the Bloch wave method in JEMS software. The error in the thickness measurements was $\pm$ 2 nm. The sample was tilted away from its zone axis to the optimum angle for EDS detection, also ensuring strong dynamical diffraction conditions are avoided. A position-averaged spectrum was taken at the centre of each precipitate over a 3 $\times$ 3 nm$^2$ area in order to remove the effect of local chemical inhomogeneity. A schematic diagram of the method is illustrated in Fig. S1 in the Supplementary Material. Starting from the Cliff-Lorimer equation,
\begin{align}
\frac{C^\mathrm{dect}_\mathrm{Ag}}{C^\mathrm{dect}_\mathrm{Al}} & =k_\mathrm{Ag-Al} \cdot \frac{I^\mathrm{dect}_\mathrm{Ag}}{I^\mathrm{dect}_\mathrm{Al}}\nonumber\\
& =\frac{C^\mathrm{prec}_\mathrm{Ag} \cdot d}{C^\mathrm{prec}_\mathrm{Al} \cdot d +(t-d)},
\label{eq:Cliff-lorimer}
\end{align}
we can deduce the composition of an embedded precipitate as follows:
\begin{equation}
C^\mathrm{prec}_\mathrm{Ag}=\frac{C^\mathrm{dect}_\mathrm{Ag} \cdot t}{d \cdot cos\theta},
\label{eq:EDS}
\end{equation}
where $C^\mathrm{prec}$ is the deduced composition of the precipitate, $C^\mathrm{dect}$ is the measured composition, $k_\mathrm{Ag-Al}$ is the Cliff-Lorimer k-factor between Ag and Al, $I^\mathrm{dect}$ is the detected characteristic intensities for quantification, $t$ is the thickness of the matrix near the precipitate, $d$ is the diameter of the precipitate and $\theta$ is the tilting angle difference between that for the thickness determination and the EDS detection. We checked the k-factor by measuring the composition of the as-water-quenched sample, as the theoretical value stored in the quantification software may easily vary by $>$10\% \cite{Williams2009}. The measured composition of 1.8 at.\% to  2.0 at.\% Ag across the sample is in good agreement with the alloy composition (Al-1.68 at.\% Ag). Although errors of the deduced compositions arising from neglecting the geometrical X-ray absorption and fluorescence remain, these errors were found not to be significant for a thin foil: according to mass-energy X-ray absorption calculations \cite{henke1993x}, about 2\% of the major characteristic X-ray for Ag ($L_\mathrm{\alpha}$=3 KeV) is absorbed by the Al matrix for a 100 nm-thick sample. Thus, the results should still be comparable between two different phases. Furthermore, the composition of one of the phases (GP zone $\epsilon$) is already known from the phase diagram \cite{mcalister1987} and many previous experiments (see for example Ref.\cite{Marquis2007}).

\subsection{STEM Image Simulations}
HAADF-STEM simulations were performed using the $\mu$STEM software \cite{Allen2015}, implementing the multislice method with quantum excitation of phonons to incorporate elastic and inelastic phonon scattering. The simulations used the optimised crystal structures of Al and $\zeta$, obtained using first principles density functional theory methods (see below). Each slice had a thickness of 1.485 \AA{} and the total sample thickness was modelled from 100 \AA{} to 600 \AA{}. Microscope parameters were matched with the experimental settings of Titan$^3$ as specified above.\par

The experimental and simulated images were analysed using ImageJ software. The brightness of the simulated images was scaled linearly to the same dynamic range as the experiment. The contrast was then adjusted to that of an experimental image by modifying the gamma correction value ($\Gamma$). The intensity was given by $I'=I^\mathrm{\Gamma}$, where $I'$ is the output intensity, and $I$ is the input intensity. Except the brightness and contrast adjustments, no other image manipulation were performed.\par

\subsection{Lattice displacements relative to Al matrix: calculations and mapping}
Geometric phase analysis (GPA) filters a lattice image according to the peaks in its fast Fourier transform (FFT) and compares that image to a reference lattice to resolve local strain in real space \cite{Hytch1998}. GPA has demonstrated an excellent spatial accuracy in agreement with classical strain theory \cite{Hytch2003,Johnson2004}. In this study, as the $\zeta$ precipitates have a super lattice of FCC aluminium and are coherently embedded within the matrix, the value calculated by GPA reflects the lattice displacements relative to FCC Al for both the precipitate phase and the matrix. An experimental STEM image with a pixel size of 1024 $\times$ 1024 was used as input, where Al matrix away from the precipitate in the same image was used as the reference. The theoretical values of lattice displacements for bulk $\zeta$ phase were calculated by comparing the the DFT-optimised structure of $\zeta$ (see details in DFT methods) in reference to the DFT-optimised Al lattice parameter using elastic strain calculation \cite{Stukowski2012} in Ovito software \cite{Stukowski2010}. The lattice parameter values of aluminium obtained from both experimental measurements ($4.04 \pm 0.05$ \AA) and DFT optimisation (4.05 \AA), as the reference for both calculations, were in good agreement. The linear scanning distortion was corrected with a standard gold cross-grating sample before imaging for geometric phase analysis. Because STEM moves the probe in a raster, scanning artefacts arise due to the time delay between measurements and accumulated error in probe position \cite{Ophus_Correcting_2016}. The noise, usually non-linear in nature due to the external field, is more predominant in the slow scanning direction comparing to the fast scanning direction. The non-linear scanning distortion was corrected with image pairs in orthogonal scan directions using the algorithm described in Ref.~\cite{Ophus_Correcting_2016}. The uncertainties in our GPA results were better than $\pm1\%$ after the distortion correction. Note that the atomic size difference between Al and Ag is negligible (about 0.5\%). Therefore, our geometric phase analysis is not sensitive to the composition. However, it is sensitive to any structural change larger than 1\%. All the calculated images were colourised with the same scale from -7\% (contraction) to 7\% (extension) for visualisation.\par

\subsection{Density Functional Theory Calculations}
First-principles density functional theory (DFT) calculations were performed using the Vienna {\em Ab initio} Simulation Package (VASP) \cite{kresse1996efficient} using the generalised gradient approximation of Perdew, Burke, and Ernzerhof (GGA-PBE) \cite{perdew1996generalized} with projector augmented wave potentials \cite{Blochl1994,kresse1999ultrasoft}. Geometrical relaxations were performed to optimise the supercells until Hellmann-Feynman forces were less than 0.01 eV/\AA, where all lattice parameters and all internal coordinates were optimised if not stated otherwise. The convergence of the relevant energy differences with respect to energy cut-off, k-point sampling and supercell size was better than 1 meV/atom.\par

The formation energies of different phases are given relative to the energy of FCC Al and Ag in the ground state. The defect energy of Ag in solid solution was calculated by an isolated Ag substitutional point defect in an Al supercell containing 108 atoms, giving a substitutional defect energy of 0.09 eV, in reasonable agreement with previous calculations of 0.02 eV using the local density approximation (LDA) \cite{Wolverton2006}. The formation energies of Ag clusters, including di-atom clusters and tri-atom clusters, were also calculated using a 108-atom Al supercell with Ag substitutions in different configurations. Planar Ag-Al structures were modelled using tetragonal or trigonal supercells in which the precipitates were surrounded above and below by Al (representing the infinitely wide two-dimensionally coherent Ag plane(s) surrounded by Al matrix), containing the equivalent of 20-24 atomic planes ($\{001\}_\mathrm{Al}$, $\{110\}_\mathrm{Al}$ or $\{111\}_\mathrm{Al}$). Sufficient numbers of Al atomic layers were used to simulate the effect of an infinitely large Al matrix. The $\zeta$ phase was investigated by assuming each bilayer was pure Ag or Al, which resulted in a composition of AgAl. The embedded $\zeta$ phase was calculated by the sandwiched structure of AgAl with the Al matrix using the supercell as described above. The $\gamma^\prime$ phase was calculated using the model by Neumann \cite{neumann1966} with lattice parameters constrained to experimental measurements \cite{rosalie2014chemical}. The bulk phases of Al, Ag, $\zeta$ (AgAl) and $\gamma$ (\ce{Ag2Al}) were fully optimised; their calculated lattice parameters were in good agreement with experiments \cite{neumann1966,howe1987} and their formation energies were consistent with previous calculations \cite{Zarkevich2002,Wolverton2006}.






\section{Results}

\subsection{Atomic Structure: HAADF-STEM Imaging, Simulation and Analysis}

\begin{figure}
\centering
\includegraphics[width=0.8\textwidth]{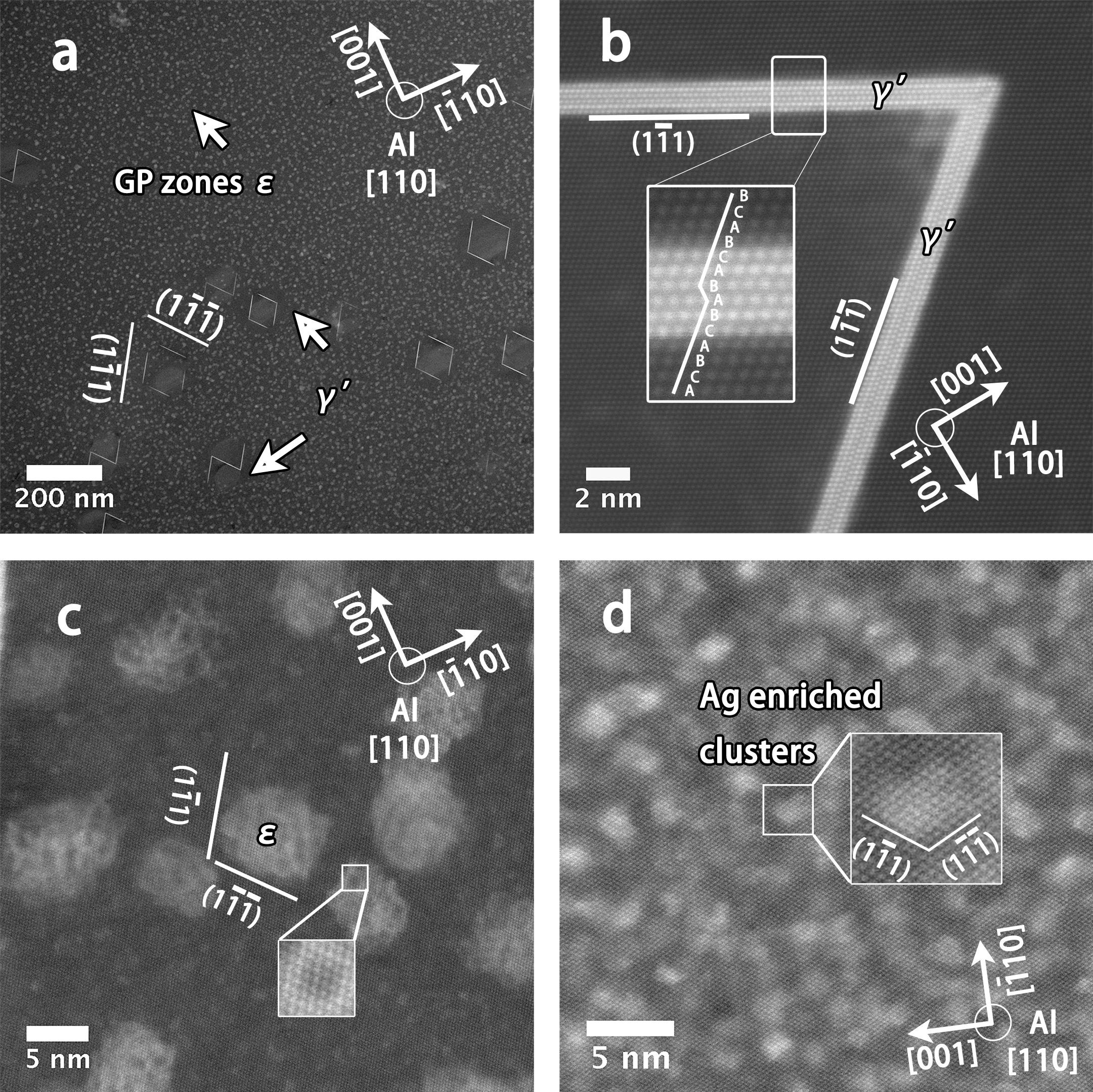}
 \caption{HAADF-STEM images of the typical microstructure for Al-1.68 at.\% Ag aged 2 h at 200\degree C after water quenching. (a) Low magnification image of $\gamma^\prime$ precipitate assemblies and GP zones; (b) high magnification image showing a $\gamma^\prime$ precipitate with the enlarged image illustrating the characteristic stacking fault associated with a HCP precipitate (ABAB stacking) embedded within the FCC matrix (ABCABC stacking); (c) high magnification images showing the $\epsilon$ GP zones. The enlarged images show the Ag depletion area inside a GP zone. (d) High magnification image of small Ag clusters formed in the as-water-quenched state. The electron beam is parallel to $\langle 110 \rangle_\mathrm{Al}$.}
 \label{fig:overview}
\end{figure}

An Al-1.68\%Ag alloy aged at 200\degree C for times varying from as-quenched to 7 days exhibits the  microstructure expected from previous studies \cite{rosalie2009,rosalie2011}: finely distributed Ag-enriched coherent precipitates known as GP zones and sparsely distributed $\gamma^\prime$ precipitates. Fig.~\ref{fig:overview}(a) shows a typical view of the alloy quenched in water and aged at 200\degree C for 2 h. The $\gamma^\prime$ precipitates are in the shape of structured assemblies, in agreement with previous findings \cite{rosalie2009}. Fig.~\ref{fig:overview}(b) confirms that $\gamma^\prime$ precipitates have an ABAB stacking embedded in the FCC aluminium matrix (ABCABC stacking). The $\gamma^\prime$ precipitates also display the expected orientation relationship of $\{111\}_\mathrm{Al}\parallel\{0001\}_\mathrm{\gamma'}$ and $\langle 110 \rangle_\mathrm{Al}\parallel\langle 11\bar{2}0 \rangle_\mathrm{\gamma'}$ with an exceptionally good lattice matching with aluminium \cite{howe1987,zakharova1966identification}. Fig.~\ref{fig:overview}(c) shows GP zones $\epsilon$ formed above the $\eta$-$\epsilon$ transition temperature displaying chemical inhomogeneity within the precipitates, where the darker columns correspond to Ag depletion. This observation is consistent with earlier X-ray \cite{dubey1991} and STEM results \cite{erni2003a} that show Ag depletes in the core and enriches in the shell. However, the detailed structure of $\epsilon$ can be considered as a multiple core-shell complex, rather than the simple model with one core as proposed previously \cite{dubey1991}. These GP zones $\epsilon$ are likely the growth product of small Ag clusters. Fig.~\ref{fig:overview}(d) shows that small Ag enriched clusters with few atoms readily exist in the as-quenched state due to decomposition, in agreement with work published over 50 years ago \cite{baur1962}.\par

\begin{figure}
\centering
 \includegraphics[width=\textwidth]{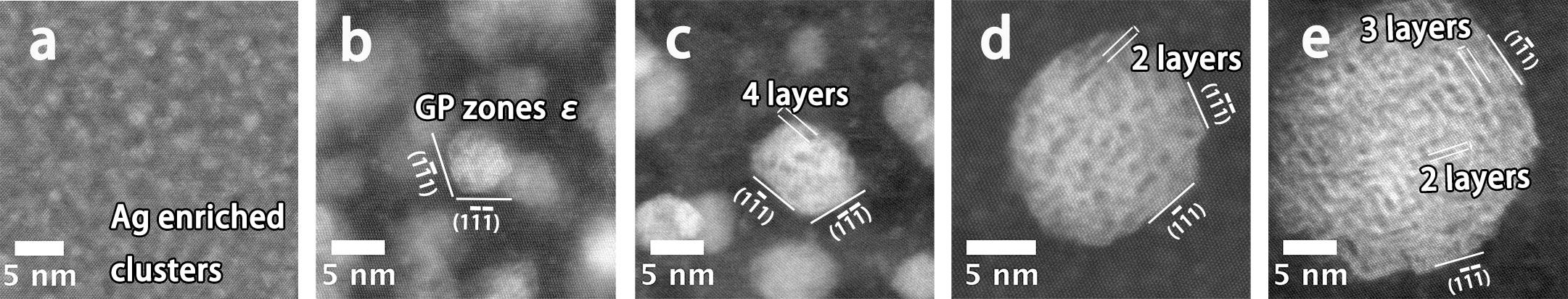}
 \caption{HAADF-STEM images of GP zones aged at 200\degree C after quenching in oil for various ageing times: (a) as-oil-quenched (b) 1 h; (c) 2 h; (d) 24 h; (e) 7 days. The ordering of Ag solute becomes increasingly defined on the $\{111\}_\mathrm{Al}$ planes while the width of the Ag depletion remains about two to four $\{111\}_\mathrm{Al}$ layers. The electron beam is parallel to $\langle 110 \rangle_\mathrm{Al}$.}
 \label{fig:GP_evolution}
\end{figure}

The chemical inhomogeneities within GP zones $\epsilon$ develops gradually by diffusion of Ag in Al. Fig.~\ref{fig:GP_evolution}(a-c) shows $\epsilon$ GP zones are more homogeneous at the early stage in the oil quenched sample. Fig.~\ref{fig:GP_evolution}(c-e) shows that Ag on $\{111\}_\mathrm{Al}$ planes becomes increasingly ordered, while the widths of the Ag depletion regions remains relatively constant at about two to four $\{111\}_\mathrm{Al}$ layers. This unique behaviour will be explained with first-principles calculations in a later section. The purpose of oil quenching was to reduce the quenched-in vacancy concentration to suppress the formation of $\gamma^\prime$ and preserve the growth of GP zones (see the different precipitation behaviours between water quenched and oil quenched samples in Fig. S2 in the Supplementary Material). This process enabled us to obtain large GP zones with diameters up to 25 nm (see Fig.~\ref{fig:GP_evolution}(e)) that were subsequently used in the following {\em in situ} annealing in the TEM.\par

\begin{figure}
\centering
 \includegraphics[width=\textwidth]{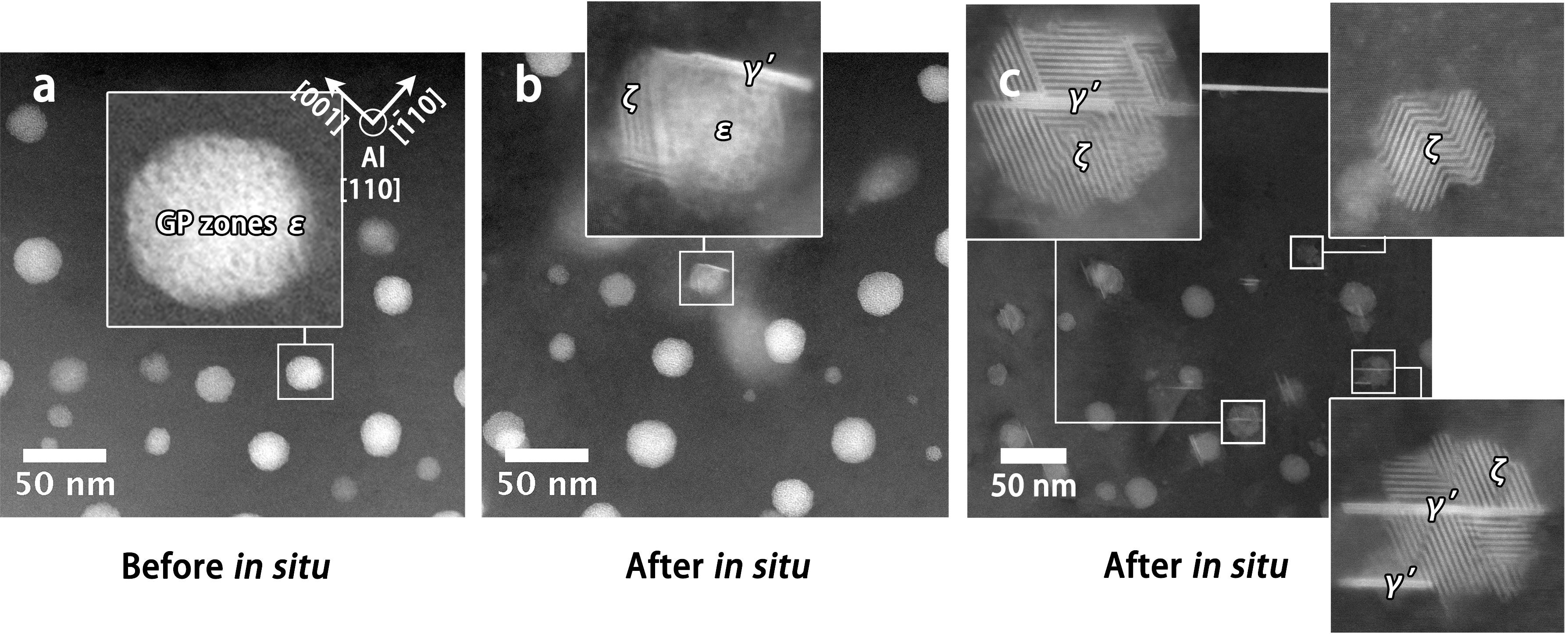}
 \caption{HAADF-STEM images of the microstructure before and after {\em in situ} annealing at 200\degree C for 3 min. The original sample is oil quenched and aged at 200\degree C for 7 days. (a) Before {\em in situ} annealing, where the enlarged image shows the GP zone before transformation; (b) after {\em in situ} annealing, where the enlarged image shows the GP zone  shown in (a) now containing $\gamma^\prime$ precipitates; (c) after {\em in situ} annealing in a different area from (a), where the enlarged images show more transformed GP zones with a layered structure. The electron beam is parallel to $\langle 110 \rangle_\mathrm{Al}$.}
 \label{fig:insitu}
\end{figure}

Fig.~\ref{fig:insitu} shows the microstructural change of the alloy containing large GP zones after 7 days ageing at 200\degree C before and after the secondary ageing within the electron microscope. A transformation occurred within a GP zone $\epsilon$ after a short time (3 min) annealing at 200\degree C inside the microscope, as shown in Fig.~\ref{fig:insitu}(a-b). This transformation did not occur for every GP zone, and the density was not uniform across the sample. Fig.~\ref{fig:insitu}(c) shows a different area of the same sample with a higher density of the transformed GP zones and a clear layered structure inside (see insets). In addition, Fig. S3 in the Supplementary Material shows the microstructure evolution during {\em in situ} annealing at 150\degree C. The layered structure originated from the local ordering of $\epsilon$. Later, $\gamma^\prime$ assemblies nucleated and grew beside the layered structure. A time-resolved phase transformation movie was recorded with an interval of 28 s between each frame (SM\_movie\_1). We tested different {\em in situ} annealing temperatures (100\degree C, 150\degree C and 200\degree C) and samples with different GP zones sizes. The results indicate that the transformation needs large GP zones and a relatively high annealing temperature ($\geq$ 150\degree C). Interestingly, Fig. S4 in the Supplementary Material shows that small GP zones actually shrunk during {\em in situ} annealing, as a result of Ag diffusion to the sample surface. In order to examine the potential effect of electron irradiation on phase transformations, we performed {\em in situ} annealing experiments without the electron beam. Results show that, without the interaction with the electron beam, newly formed layered structure and $\gamma^\prime$ precipitates were still found (see Fig. S5 in the Supplementary Material).\par

Fig.~\ref{fig:Z}(a) shows that the ordered phase is clearly distinct from GP zones $\epsilon$ or $\gamma^\prime$ plates. We named this new phase as $\zeta$. Specifically, $\gamma^\prime$ has an ABAB stacking while $\zeta$ follows the ABCABC stacking of FCC Al. The phase exhibits different domains corresponding to different $\{111\}_\mathrm{Al}$ variants, as viewed along a $\langle 110 \rangle_\mathrm{Al}$ direction (see Fig.~\ref{fig:Z}(a)). Fig.~\ref{fig:Z}(b) shows that Ag can be depleted in some atomic columns within the Ag-enriched bilayers, as viewed along $\langle 112 \rangle_\mathrm{Al}$. But the depletion has no periodicity, and the overall intensity is quite uniform. After imaging in the $\langle 110 \rangle_\mathrm{Al}$ and $\langle 112 \rangle_\mathrm{Al}$ directions, we conclude that the new phase has a super lattice structure of FCC Al and consists of alternative Ag enriched bilayers and Al enriched bilayers on $\{111\}_\mathrm{Al}$ planes.

In many cases, $\gamma^\prime$ precipitates formed inside GP zones $\epsilon$ and introduced Ag depletion at their coherent interfaces ($\{111\}_\mathrm{Al/\epsilon} \parallel \{0001\}_\mathrm{\gamma'}$). As shown in Fig.~\ref{fig:Z}(c), the widths of the Ag depletion gaps are about two to three atomic layers. Away from the coherent interfaces of $\gamma^\prime$ precipitates, Ag enriches and then depletes again in a specific frequency that is similar to the modulation of $\zeta$, suggesting these regions are poorly ordered version of $\zeta$ phase. The precipitates, $\zeta$ and $\gamma^\prime$, transformed from GP zones, can be easily found on a large scale after {\em in situ} annealing (see Fig.~\ref{fig:insitu}(c)). However, these structures are rarely found in samples obtained using conventional heat treatment with one example shown in Fig.~\ref{fig:Z}(d). This sample was water quenched and aged at 200\degree C for 2 h without {\em in situ} annealing. The original GP zone developed Ag depletion inside and formed $\gamma^\prime$ and $\zeta$ at its edges. Interestingly, $\gamma^\prime$ precipitates also induced the ordering of Ag at their coherent interfaces similar to Fig.~\ref{fig:Z}(c) (as indicated by arrows).\par

\begin{figure}
\centering
 \includegraphics[width=\textwidth]{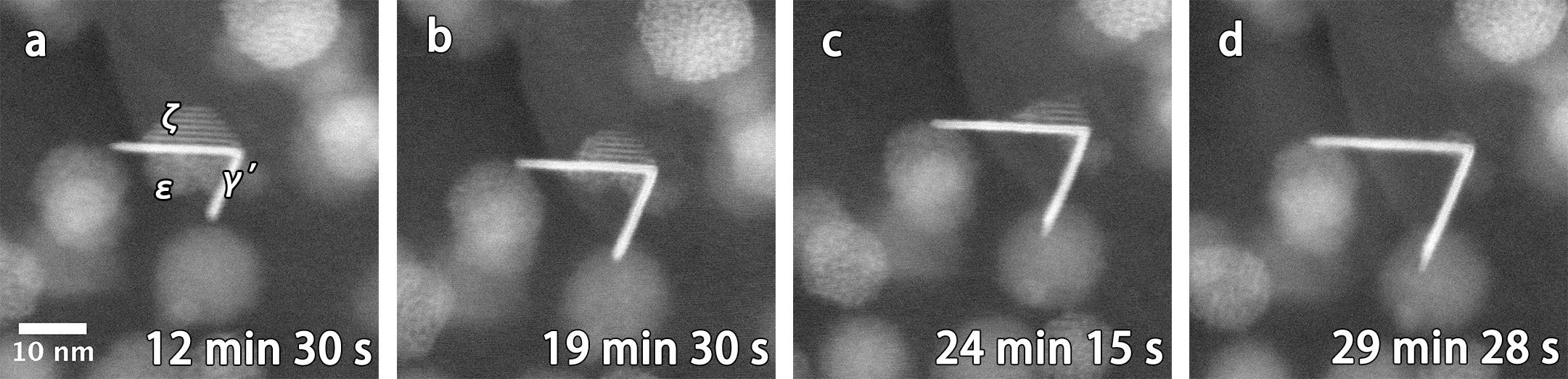}
 \caption{HAADF-STEM images of evolution of $\zeta$ to $\gamma^\prime$ during {\em in situ} annealing at 200\degree C for times as labelled. The original sample was oil quenched and aged at 200\degree C for 24 h. The electron beam is parallel to $\langle 110 \rangle_\mathrm{Al}$.}
 \label{fig:Z_to_G}
\end{figure}

The metastable $\zeta$ phase will eventually transform into HCP $\gamma^\prime$. This is evident by Fig.~\ref{fig:Z_to_G} where the transformation during {\em in situ} annealing was recorded. Fig.~\ref{fig:Z_to_G}(a) shows a GP zone in which both $\zeta$ and $\gamma^\prime$ formed inside. Fig.~\ref{fig:Z_to_G}(b-d) shows $\zeta$ and the remaining GP zone shrinking gradually with increasing ageing time until their full dissolution, to the benefit of the growing $\gamma^\prime$ precipitates. This suggests that Ag atoms diffuse from the metastable phases to the more stable HCP phase.

\begin{figure}
\centering
 \includegraphics[width=\textwidth]{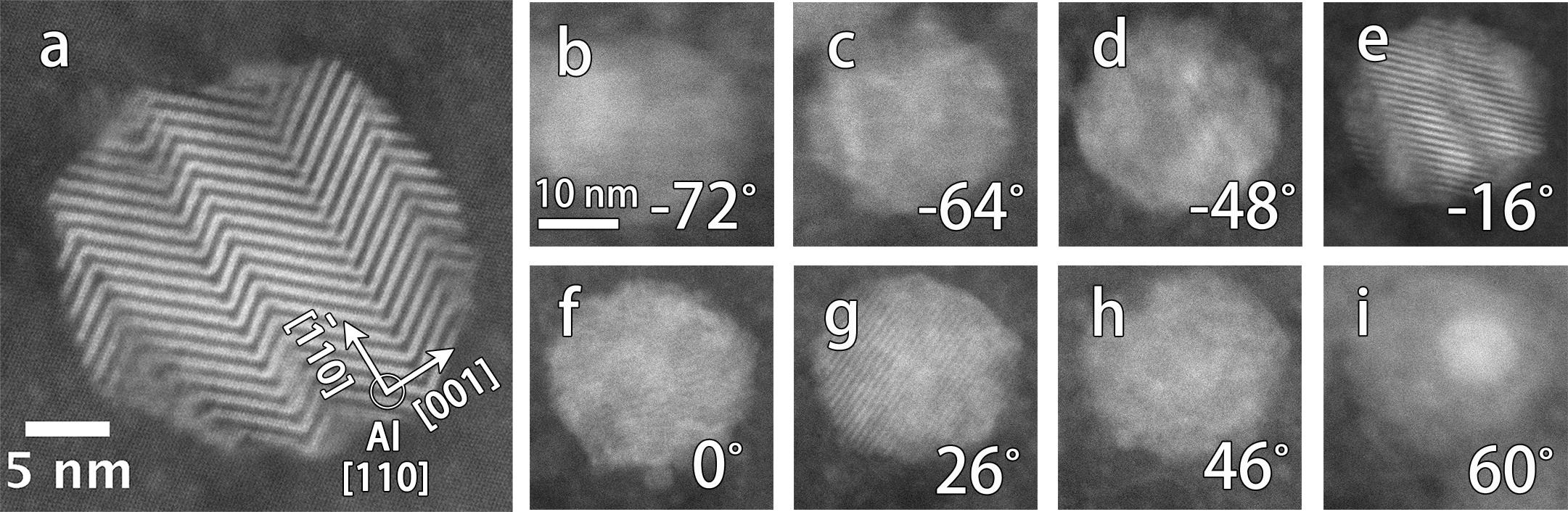}
 \caption{(a) HAADF-STEM images of a bi-layered phase viewed along the $\langle 110 \rangle_\mathrm{Al}$ zone axis, showing domains formed by two variants of $\{111\}_\mathrm{Al}$ bi-layers. (b-i) HAADF-STEM tilt series of the same bi-layered precipitate phase for a tilt range from -73\degree{} to 64\degree. The tilt angles are as labelled in each image.}
 \label{fig:tilt-series}
\end{figure}

Precipitation of $\zeta$ accompanies $\gamma^\prime$ in most cases, but some isolated examples of $\zeta$ without $\gamma^\prime$ were also found. A tilt series was performed for a given $\zeta$ precipitate as shown in Fig.~\ref{fig:tilt-series}. Fig.~\ref{fig:tilt-series}(a) shows a $\zeta$ precipitate viewed along the $\langle 110 \rangle_\mathrm{Al}$ zone axis with different domains of $\{111\}_\mathrm{Al}$ bilayers and their domain boundaries on \{001\} planes. Fig.~\ref{fig:tilt-series}(b-i) shows the tilt series of the same precipitate in an angular range of -73\degree{} to 64\degree, where we performed 2\degree{} per tilt under $\pm 60\degree{}$ and 1\degree{} per tilt above $\pm 60\degree{}$. The layered contrast can be seen in Fig.~\ref{fig:tilt-series}(e) and (g), corresponding to the $\{111\}_\mathrm{Al}$ bilayered variants. Throughout the tilt series, no $\gamma^\prime$ precipitate is visible. The drift-corrected tilt series can be found in the Supplementary Material (SM\_movie\_2), showing the structural difference between the $\zeta$ phase and surrounding GP zones $\epsilon$ during tilting. \par

It is important to address the chemical composition and crystal structure of $\zeta$. However, determining the composition of each layer is challenging for an embedded precipitate, particularly having matrix above and below the precipitate in the electron beam direction. Thus, we propose the simplest model in which each layer is pure Al or Ag and examine the validity of the model in terms of EDS analysis, HAADF-STEM image simulations, atomic positions and energetics. We examined 8 different GP zones $\epsilon$ and 10 different $\zeta$ precipitates from 3 different grains with sizes ranging from 10 nm to 25 nm and the thickness of each surrounding matrix ranging from 20 nm to 100 nm with two different EDS systems. The deduced compositions are essentially the same for GP zone $\epsilon$ (38 at.\% Ag at 200\degree C) and $\zeta$ (40 at.\% Ag at 200\degree C) with a small standard deviation between different datasets (3 at.\% Ag for each phase). The uncertainties are 6 at.\% Ag for each phase when quantified spectrums using the Cliff-Lorimer ratio method. Notably, the measured  $\epsilon$ composition is in excellent agreement with previous X-ray results (38 at.\% Ag at 200\degree C) \cite{dubey1991} and atom probe tomography (40 at.\% Ag at 200\degree C) \cite{Marquis2007}. The compositional analysis strongly suggests that the $\epsilon$-$\zeta$ transformation involves a minimal chemical change, if any, and only rearrangement of the solute atoms within GP zones.

\begin{figure*}[htbp]
\centering
\includegraphics[width=1\linewidth]{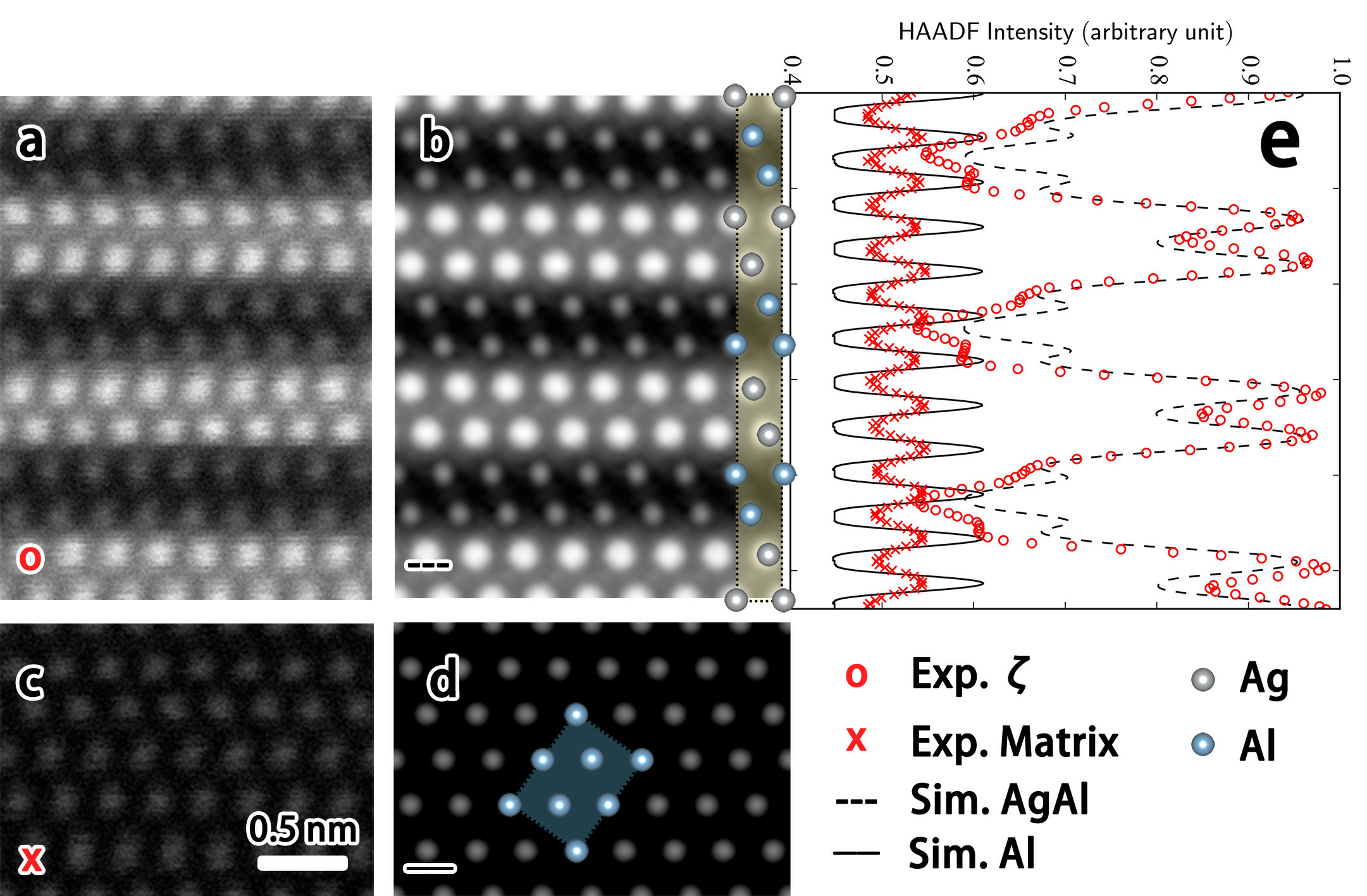}
\caption{Atomic-resolution HAADF-STEM images obtained from experiments and simulations for a $\zeta$ precipitate: (a) raw image section of embedded $\zeta$ precipitate; (b) simulated image of bi-layered AgAl [thickness: 30 nm] with the atomic structure overlaid [grey: Ag, blue: Al]; (c) raw image section of the matrix near the embedded $\zeta$ shown in (a); simulated image of Al [thickness: 30 nm] with the atomic structure overlaid. (e) intensity profile of the experimental and simulated images in $\zeta$ compared to that of the Al matrix. The orientation of the intensity profile is aligned with images (a) and (c). The electron beam is parallel to $\langle 110 \rangle_\mathrm{Al}$.}
\label{fig:STEM_sim}
\end{figure*}

Fig.~\ref{fig:STEM_sim} shows the embedded $\zeta$ phase and nearby Al matrix from the HAADF-STEM image, where both regions had approximately the same thickness. Therefore, the structures were considered as $\zeta$ and Al with the same thickness sandwiched by Al matrix above and below the phases along the beam direction. The experimental images were compared with the simulated images for $\zeta$ (AgAl) and Al without matrix. The peak positions of Al or $\zeta$ (AgAl) determined in the experiments and simulations match reasonably well. The shoulders in the experimental intensity profile of $\zeta$ phase correspond to Al columns in the AgAl model. In the experimental images, the Al-enriched columns in $\zeta$ are brighter than the Al matrix. However, it does not necessarily mean there is Ag within those columns, as the simulated images also show the same phenomenon. This means that the recorded intensities corresponding to Al columns actually contains a contribution from the scattering by neighbouring Ag columns. It is hard to preclude the presence of Ag in the Al-enriched columns, but there is a distinct possibility that beam spreading \cite{Dwyer_Scattering_2003} causes the increase in intensity. The matrix above and below the precipitate would lower the contrast between the Ag-enriched and Al-enriched layers compared to the model. Besides, the atomic positions of bulk Al differs from that of $\zeta$ in both experiments and simulations, which means  that matrix above and below the precipitate might  blur the HAADF-STEM image of atomic columns for an embedded $\zeta$. The $\mu$STEM algorithm takes no account of source size that also significantly blurs any experimental STEM image. But the HAADF-STEM intensity is dominated by Ag, and hence those effects should not change the validity of our results. \par

\begin{figure*}[htbp]
\centering
\includegraphics[width=1\linewidth]{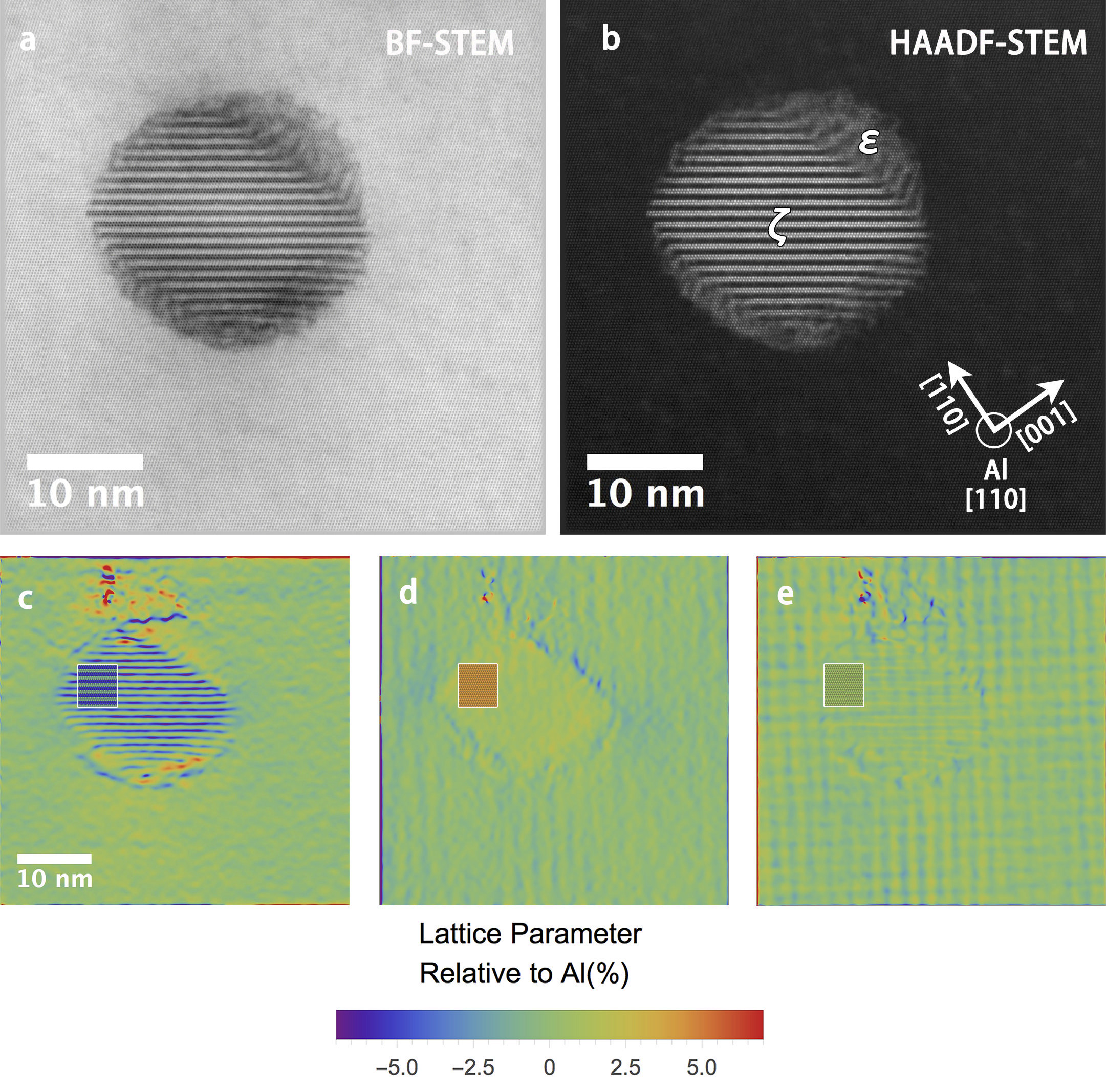}
\caption{Lattice displacements mapping of a $\zeta$ precipitate in aluminium. Original (a) bright field (BF)-STEM and (b) HAADF-STEM images for geometric phase analysis (GPA). The GPA results were compared with simulations based on the DFT-optimised structure of the bi-layered AgAl model (in the white box) in the following directions (c) normal, (d) parallel and (e) sheared with respect to the $\zeta$ basal planes.}
\label{fig:Strain_analysis}
\end{figure*}

The modulation in the chemical composition by 4 (2 Ag and 2 Al) and the stacking ordering by 3 (ABCABC) require at least 12 $\{111\}_\mathrm{Al}$ planes to achieve the periodicity of $\zeta$, as shown in the atomic structure in Fig.~\ref{fig:STEM_sim}(b). The bi-layered AgAl model of $\zeta$ is a trigonal crystal with a space group of $R\bar{3}m$ (hexagonal axes). The lattice parameters for the bulk $\zeta$ phase are $a_\mathrm{DFT}$=2.97 \AA{} and $c_\mathrm{DFT}$=26.88 \AA{} after DFT optimisation, which agree reasonably well with the experimental measurements for embedded $\zeta$ precipitates of $a_\mathrm{exp}$=$2.88 \pm 0.05$ \AA{} and $c_\mathrm{exp}$=$27.35 \pm 0.05$ \AA. The embedded $\zeta$ precipitate calculation gives a much better match with $a_\mathrm{DFT}^{emb}$=2.92 \AA{} and $c_\mathrm{DFT}^{emb}$=27.26 \AA{}, which means $\zeta$ precipitates are deformed to accommodate the change in lattice parameters compared with Al. When embedded within the Al matrix, $\zeta$ is coherent with the matrix with an orientation relationship of $\{111\}_\mathrm{Al}\parallel\{001\}_\mathrm{\zeta}$ and $\langle 110 \rangle_\mathrm{Al}\parallel\langle 100 \rangle_\mathrm{\zeta}$. Table~\ref{table:site} lists the coordinates of Ag- and Al-containing sites in $\zeta$, showing an exceptionally good agreement between experiments and calculations. Close inspection of the lattice sites reveals that the spacings of the basal planes (including Ag-Ag, Ag-Al and Al-Al) vary along $\langle 001 \rangle_\mathrm{\zeta}\parallel\langle 111 \rangle_\mathrm{Al}$. This is further demonstrated in Fig.~\ref{fig:Strain_analysis} using geometric phase analysis (GPA) of a distortion-corrected HAADF-STEM image in order to map these lattice displacements. Theoretical displacements of $\zeta$ relative to Al were calculated based on the DFT-optimised structure of the AgAl model and compared with GPA results in each direction. The effect of scanning noise is demonstrated in  Fig. S6 in the Supplementary Material. Fig.~\ref{fig:Strain_analysis}(c) shows the displacements in the direction normal to $\zeta$ basal planes have a clear modulation in both the GPA result and the DFT-optimised structure. Within the $\zeta$ phase, the local contraction of the lattice is significant at the Al sites as deduced from both GPA and DFT (GPA: -6.5\% and DFT: -6.6\%), but less at the Ag sites (GPA: 0\% and DFT: -1.9\%). This remarkable lattice variation between the bilayers of a $\zeta$ precipitate is not due to the atomic size difference between Al and Ag; instead, it is a result of different spacings of the basal planes in $\zeta$ phase. Specifically, the interplanar distance between Al-Ag in $\zeta$ phase is greatly decreased to 2.26 \AA{} compared to the spacing of 2.34 \AA{} between $\{111\}_\mathrm{Al}$ planes, as shown in Fig.~\ref{fig:STEM_sim}. The chemical composition has to be significantly different between the sequential bilayers to cause a change of the bond length, which also demonstrate the validity of our model regarding the atomic positions. The displacements are small in the direction along the basal planes of $\zeta$ (GPA: 0.8\% and DFT: 1.8\%) in Fig.~\ref{fig:Strain_analysis}(d) and negligible in the shear direction in Fig.~\ref{fig:Strain_analysis}(e), which also represents good agreement between GPA and DFT. Finally, the $\zeta$ precipitate is coherent within the matrix without any misfit dislocation as evident from the BF-STEM image (see Fig.~\ref{fig:Strain_analysis}(a)). The remaining GP zone $\epsilon$ (as labelled in Fig.~\ref{fig:Strain_analysis}(b)) is almost strain-free in all directions, as shown in Fig.~\ref{fig:Strain_analysis}(c-e). \par

\subsection{Energetics: First-principles Calculations, Strain Energy and Entropy}
The clustering process during the decomposition of the solid solution is governed by the energy of different solute configurations and the barriers between them. To understand the clustering of Ag in Al, we calculated the formation energy of various Ag clusters by DFT as shown in Table~\ref{table:DFT}. The solid solution is not energetically stable with a defect energy of 89 meV/Ag atom, which drives the solid solution to decompose. For bi-atom Ag clusters, the nearest neighbours along $\langle 110 \rangle_\mathrm{Al}$ are preferred compared with the second nearest neighbours along $\langle 001 \rangle_\mathrm{Al}$. A tri-atom cluster on either $\{110\}_\mathrm{Al}$ or $\{111\}_\mathrm{Al}$ planes is almost as stable as segregated Al and Ag in bulk. The calculated formation energy of Ag monolayer aggregation on $\{111\}_\mathrm{Al}$ is -65 meV/Ag atom, which is significantly more stable than Ag on the other low-index crystallography planes. This energy is substantial, given the thermal energy at 200\degree C is 40 meV. Not surprisingly, $\{111\}_\mathrm{Al}$ planes become the basal planes for $\zeta$ and the HCP phases $\gamma^\prime$/$\gamma$.\par

\begin{figure}
\centering
  \includegraphics[width=0.5\textwidth]{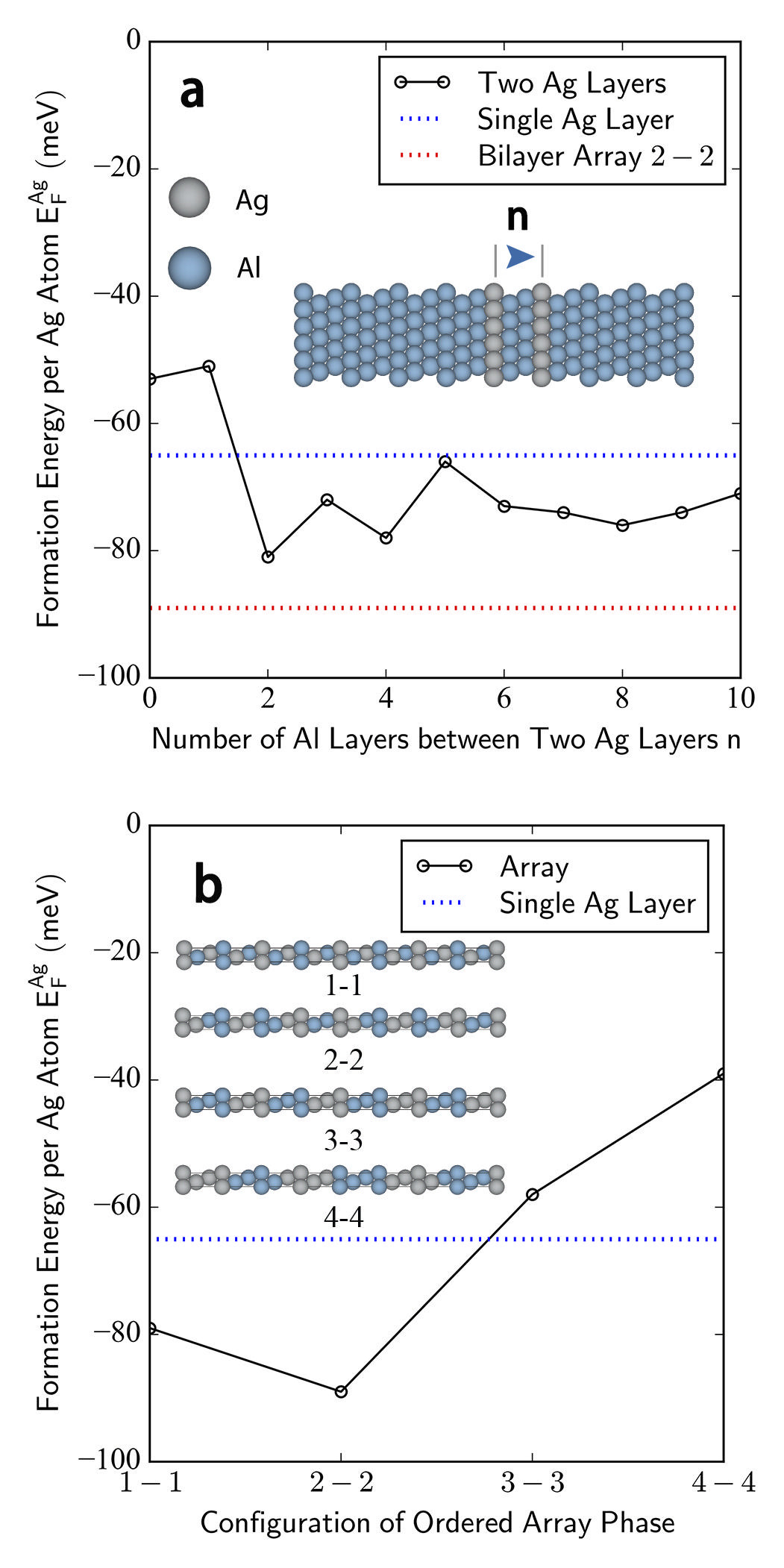}
 \caption{DFT calculations illustrating the preference of $\{111\}_\mathrm{Al}$ planes for Ag aggregation in aluminium. (a) The energetics of two Ag $\{111\}_\mathrm{Al}$ planes separated by a varying number ``n'' of Al planes. For instance, ``2'' means there are two Al atomic layers between two Ag layers as shown in the schematic diagram. (b) Energetics of different periodic arrays with a composition of AgAl. For instance, ``1-1'' means the modulation of one Ag layer and one Al layer as shown in the schematic diagram.}
 \label{fig:Ag111}
\end{figure}

\begin{figure}
\centering
  \includegraphics[width=0.5\textwidth]{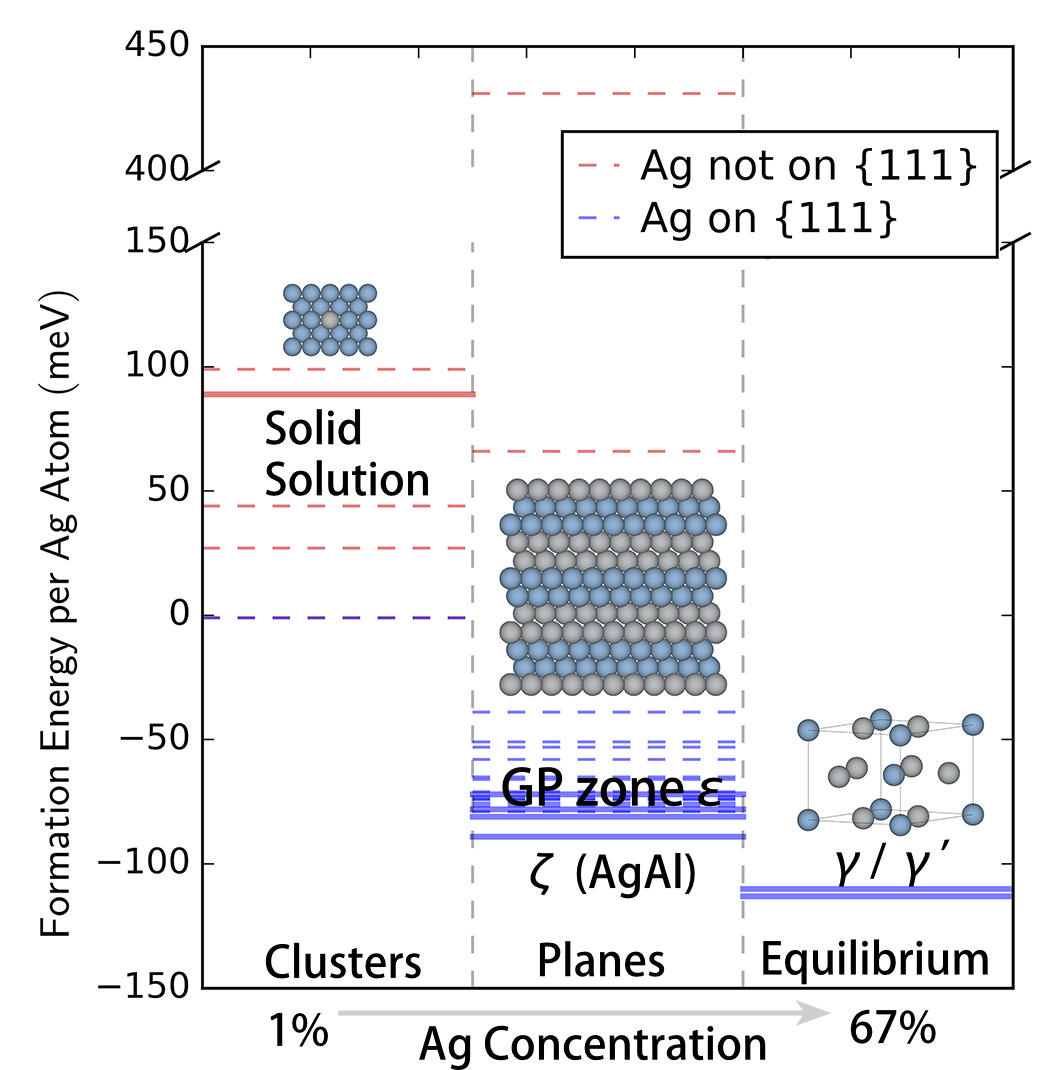}
 \caption{Energetics of Ag clustering from the solid solution to the equilibrium $\gamma$ phase. Different configurations of Ag on $\{111\}_\mathrm{Al}$ planes are shown in blue while Ag clustering on other crystallographic planes are shown in red. The phases in the transformation sequence are highlighted with a bold unbroken line with their corresponding names and atomic structures, while other configurations calculated are shown as dashed lines.}
 \label{fig:DFT_clustering}
\end{figure}

With Ag placed on $\{111\}_\mathrm{Al}$ planes, we investigated the preferential distance between two Ag layers in aluminium. A series of calculations were performed with varying Al layers between two Ag layers as shown in Fig.~\ref{fig:Ag111}. Interestingly, a distance range of two to four $\{111\}_\mathrm{Al}$ Al layers between the two Ag layers is favourable, as a closer spacing yields a considerably higher energy state. The lowest energy structure corresponds to two Al layers between two Ag layers. The energy is further lowered when Ag layers are assembled according to a periodic layered array as shown in Fig.~\ref{fig:Ag111}(b). For a fixed composition of AgAl and ABCABC stacking, the bi-layered array is the most stable, which demonstrates the validity of our model for the $\zeta$ phase from an energetics perspective. As summarised in Fig.~\ref{fig:DFT_clustering}, each phase transformation is accompanied by a decrease in energy. From clusters containing only a few atoms to the equilibrium phase $\gamma$, the Al-Ag system lowers its energy by ordering Ag solute on $\{111\}_\mathrm{Al}$ planes in the Al matrix. The formation energy of the complex GP zone $\epsilon$ is about 72-81 meV/Ag atom, as approximated by the energy range calculated for layered Ag aggregation with favourable spacings in aluminium. The formation energy of the new phase $\zeta$ (AgAl) is 89 meV/Ag atom, the lowest in terms of ordered Ag planes on $\{111\}_\mathrm{Al}$ prior to the FCC-HCP transformation, which agrees with our {\em in situ} observations (see Fig.~\ref{fig:Z_to_G}).\par

The enthalpy difference between the GP zone $\epsilon$ and the $\zeta$ phase is as little as 8-17 meV/Ag atom. The free energy difference between those two phases should be even smaller: the strain energy and configurational entropy do not favour the $\epsilon$-$\zeta$ transformation. Specifically, GP zones $\epsilon$ are coherent with almost no strain in the Al matrix, as evident in Fig.~\ref{fig:Strain_analysis}. But $\zeta$ precipitates are coherent with strain according to our experiments and simulations, which gives $\zeta$ precipitates a higher strain energy comparing to GP zones $\epsilon$. Based on elastic theory for a spherical precipitate with anisotropic strain, the strain energy contribution $E_\mathrm{e}$ of an embedded $\zeta$ precipitate is estimated using the following equation as
\begin{equation}
E_\mathrm{e}= \mu \delta^2 V,
\label{eq:strain}
\end{equation}
where $\mu$ is the shear modulus, which is assumed to be the same for both the $\zeta$ precipitate and aluminium matrix; $\delta$ is the averaged strain along $\langle 001 \rangle_\mathrm{\zeta}\parallel\langle 111 \rangle_\mathrm{Al}$, given the strains in other directions are negligible. Here we assume that Poisson's ratio is 1/3 for both the matrix and the precipitate. V is the atomic volume of $\zeta$, {\em i.e.} the ratio of the unit cell volume and the number of atoms within the cell. The strain energy is estimated to be 3 meV/atom, or 6 meV/Ag atom with the composition of AgAl. Also, GP zones $\epsilon$ have a higher configurational entropy due to the chemical inhomogeneities, in contrast to a well-ordered phase like $\zeta$ phase. The entropy of GP zones $\epsilon$ is hard to estimated with the complex structure, but the value should be in between of that for a well-ordered phase and an ideal mixing alloy using the equation
\begin{equation}
\Delta S_\mathrm{mix}=-k_\mathrm{B}(XlnX+(1-X)ln(1-X)),
\label{eq:entropy}
\end{equation}
where $\Delta S_\mathrm{mix}$ is the mixing entropy of the binary alloy, $k_\mathrm{B}$ is the Boltzmann constant, $X$ is the composition of the binary alloy. For GP zones $\epsilon$ with the composition of Al-40 at.\%Ag, the configurational entropy is 0.67 $k_\mathrm{B}$/atom. In practice, the Bragg-Williams approximation of Eq.~\ref{eq:entropy} overestimates the configurational entropy as it neglects any ordering. The short range and long range ordering can be incorporated into the equation by considering the probabilities of bonds between Al-Al, Ag-Ag and Al-Ag \cite{hillert1998phase}. According to thermodynamics, such probabilities can be calculated based on the bond energies, usually between nearest neighbours. The bond energies are assumed to be constant while the bond fractions are varying for different configurations, no matter whether solute atoms are isolated or clustered. This simplification violates our DFT calculations that Al-Ag bond is unstable in solid solution but it is stable when Ag atoms are placed on $\{111\}_\mathrm{Al}$ planes. An accurate 3D reconstruction of the chemical distribution from the tilt series should be useful for the direct measurement of ordering (see SM\_movie\_2). New numerical computation techniques need to be developed for the purpose of entropy estimation. Nevertheless, the contribution from configurational entropy is small during phase transformations in the Al-Ag system, that otherwise prevents any kind of ordering and phase separation. We have not considered vibrational entropy in this paper.

\section{Discussion}
The bilayer phase first reported herein is a new phase in the Al-Ag system. We propose to name it $\zeta$ phase, by analogy with the patterned skin of the zebra. HAADF-STEM images show a clear picture of Ag ordering on $\{111\}_\mathrm{Al}$ planes starting from a small cluster to a large GP zone (see Fig.~\ref{fig:GP_evolution}). The positive defect energy of Ag in aluminium is consistent with previous calculations \cite{wolverton2007} that explains the driving force for the decomposition \cite{baur1962}. For comparison, Au has almost an identical size to Ag, yet Au displays a very negative defect energy in Al \cite{Bourgeois_The_2016}. It is the electronic difference between Ag and Au in aluminium that leads to completely different clustering and precipitation behaviours, either in the binary alloys \cite{Bourgeois_The_2016}[this work] or when they are added to Al-Cu alloys \cite{Rosalie2012,Chen_The_2017}. Our DFT calculations also illustrate the preference of Ag aggregation on $\{111\}_\mathrm{Al}$ planes in Al-Ag binary alloys, which also occurs at the early stage of ageing in Al-Cu-Mg-Ag alloys \cite{hono1994pre,murayama1998three} and Al-Cu-Li-Mg-Ag alloys \cite{murayama2001}. During ageing, $\{111\}_\mathrm{Al}$ planes enriched in Ag within GP zones $\epsilon$ begin to move away from each other and form Ag depletion regions as shown in Fig.~\ref{fig:GP_evolution}. The ordering of Ag clearly increases with ageing time, while the depletion width remains about two to four $\{111\}_\mathrm{Al}$ layers. The unique clustering behaviour can be understood from our DFT calculation that Ag prefers to be on $\{111\}_\mathrm{Al}$ planes but not with the $\{111\}_\mathrm{Al}$ planes close to each other. The favourable spacing is around two to four Al $\{111\}_\mathrm{Al}$ planes, which is in excellent agreement with our experiments. The local ordering within GP zones $\epsilon$ develops faster in the water-quenched sample than the oil-quenched sample, because more quenched-in vacancies are present to mediate diffusion. Ag needs diffusion to achieve the long range ordering exhibited by the bi-layered $\zeta$ phase to further lower the energy of the system. Based on DFT alone, one cannot rationalise the difference in $\zeta$ phase formation between conventional heat treatments and {\em in situ} annealing experiments. The free energy landscape locates the transformation pathways between different phases. After considering the strain energy and the entropy contribution, there is almost no energy difference between GP zone $\epsilon$ and the $\zeta$ phase. But the rearrangement of Ag atoms associated with the $\epsilon$-$\zeta$ transformation is expected to have a high energy barrier. Therefore, the local energy minimum state of $\zeta$ is hardly visited during the precipitation in Al-Ag alloys. However, the experimental fact that GP zones $\epsilon$ transform to the $\zeta$ phase and eventually $\gamma^\prime$ phase demonstrates that the free energy of $\zeta$ phase is indeed lower than that of $\epsilon$ phase. It means that their thermal histories must be taken into account to understand different phase transformation pathways. We may appreciate this phenomenon by considering that the $\zeta$ phase evolves through the local ordering of GP zone $\epsilon$ on $\{111\}_\mathrm{Al}$. Vacancies can lower the energy barrier of  substitutional diffusion during the ordering of Ag atoms. For conventional heat treatments with water quenching and sequential ageing, a substantial amount of quenched-in vacancies are present, thus helping early stage clustering or providing defects for heterogeneous nucleation (see Fig. S2 in the Supplementary Material). Indeed, $\gamma^\prime$/ $\gamma$ assemblies nucleate at dislocation loops \cite{rosalie2009}, which bypasses intermediate phases like $\epsilon$ phase and $\zeta$ phase. The quenched-in vacancies usually run out quickly at the early stage of ageing before GP zones $\epsilon$ grow large enough to exhibit the local ordering of Ag on $\{111\}_\mathrm{Al}$. This may explain why the $\epsilon$-$\zeta$ transformation is rarely observed using conventional heat treatments, given the extensive studies on this system in the last century. Often the reaction within the thin TEM specimen differs from that in the bulk, both due to the surface effect and the electron irradiation. Electron irradiation indeed can substantially lower the energy barrier for diffusion of vacancies, as we quantitatively measured in our recent study of {\em in situ} annealing of voids in aluminium \cite{zhang2016void}. However, according to our {\em in situ} annealing experiment without electron beam, $\zeta$ and $\gamma^\prime$ were still found to form within GP zones (see Fig. S5 in the Supplementary Material). This demonstrates that electron beam irradiation is not responsible for those transformations. When considering surface effects, there is a depth dependency of the vacancy formation energy at the Al surface \cite{Gupta_Depth_2016}. In general, a vacancy has a lower formation energy at the surface than in the bulk, which leads to a vacancy flux from the surface to the bulk. Diffusion calculations using Fick's equations similar to what was used in our previous work \cite{zhang2016void} suggest that such a vacancy flux can be significant for an ultra-thin sample at a temperature higher than 100\degree C, as was the case for {\em in situ} annealing experiments. The fact that small GP zones shrink during {\em in situ} annealing is an indication of such vacancy flux (see Fig. S4 in the Supplementary Material). The induced vacancies are also likely to be the source for Shockley partial dislocations, which is required for $\gamma^\prime$ formation within GP zones $\epsilon$. The oil quenched samples with large GP zones after long ageing times are depleted of vacancies. When vacancies are induced to mediate solute diffusion, those large GP zones with the local ordering of Ag on $\{111\}_\mathrm{Al}$ act as a template for $\zeta$ formation. We will report a quantitative study on the {\em in situ} experiments to elucidate the vacancy-induced transformation in the future.

The existence of the $\zeta$ phase suggests a new phase transformation approach that gives a more gradual change in terms of the chemical compositions and atomic structures. A $\zeta$ precipitate (50 at.\% of Ag in the AgAl model) develops from the increased local ordering of a GP zone $\epsilon$ (~40 at.\% of Ag at 200\degree C) \cite{Marquis2007,mcalister1987} before transforming into HCP $\gamma^\prime$/ $\gamma$ (~67 at.\% Ag) \cite{neumann1966,howe1987}. Fig.~\ref{fig:insitu} and Fig.~\ref{fig:Z_to_G} clearly show that $\zeta$ is an intermediate phase between GP zone $\epsilon$ and $\gamma^\prime$. The absence of shear in $\zeta$ minimises the energy barrier for its formation, which is considered to restrict $\gamma^\prime$ nucleation \cite{rosalie2011}. Previous calculations also have shown that pure Ag layers lower the stacking fault energy in Al \cite{Finkenstadt2006}, which offers a pathway for a $\zeta$ to $\gamma^\prime$ transformation. However, some questions are still open regarding the relationship between $\zeta$ and $\gamma^\prime$. Although the tilt series indicates that $\zeta$ can form independently from $\gamma^\prime$ (See Fig.~\ref{fig:tilt-series}), the two metastable phases are generally seen in association with one another (Fig.~\ref{fig:insitu}(b-c),~\ref{fig:Z},~\ref{fig:Z_to_G}). The {\em in situ} movie also suggests that the formation of one phase may assist in the nucleation and/or growth of the other. However, the transformation from $\zeta$ to $\gamma^\prime$ is not understood yet at the atomistic level. We have seen $\zeta$ absorbed by an existing $\gamma^\prime$ assembly in Fig.~\ref{fig:Z_to_G}, instead of initiating new $\zeta$ precipitates. In the previous phase diagram \cite{mcalister1987}, there is no intermediate phase in the composition range between \ce{Ag2Al} and Al except the metastable GP zones. However, several possible structures with the composition of AgAl were predicted using cluster expansions of DFT results \cite{zarkevich2003}. But all the predictions were HCP structures (ABAB stacking). The large periodic size of $\zeta$ along $\langle 001 \rangle_\mathrm{\zeta}\parallel\langle 111 \rangle_\mathrm{Al}$ means such structures are difficult to predict via the cluster expansion method. This points out the importance of atomic resolution electron microscopy for providing critical structural information for atomistic calculations.\par

The $\zeta$ phase in the Al-Ag system has structural similarities with layered Ag segregations to precipitate interfaces in various aluminium alloys. It is very interesting that Al-Ag alloys have a poor mechanical performance but numerous aluminium alloys with a minor addition of Ag constitute the strongest and most thermally stable series \cite{Polmear2005}. Taking the famous example of Al-Cu-Mg-Ag alloys, the $\Omega$ phase is responsible for their outstanding mechanical performance and thermal stability \cite{Polmear2005}. The $\Omega$ phase is considered as a distorted $\theta$ (\ce{Al2Cu}) on $\{111\}_\mathrm{Al}$, which is originally body-centred tetragonal forming on $\{100\}_\mathrm{Al}$ planes \cite{Auld1986,Knowles1998}. To reorient Cu atoms from $\{100\}_\mathrm{Al}$ planes to $\{111\}_\mathrm{Al}$ planes, Mg is essential to minimise the misfit of the $\Omega$ phase along its c-axis, which can be as large as -9.3\% matching half unit cell of the $\Omega$ phase with multiples of $\{111\}_\mathrm{Al}$ d-spacing \cite{Kang2014}. Indeed, the $\Omega$ phase is not found in Al-Cu or Al-Cu-Ag alloys \cite{Rosalie2012}, and only very few $\Omega$ precipitates appear in Al-Cu-Mg alloys where the dominant precipitate phase is the S phase \cite{Polmear2005,Sha2010}. However, this fact as well as our unpublished DFT calculations \cite{Zhang_database_future} suggest Mg itself does not have much tendency to drive the segregation of Cu atoms on $\{111\}_\mathrm{Al}$. As we have shown above that Ag prefers to aggregate on $\{111\}_\mathrm{Al}$ planes in aluminium (see Table~\ref{table:DFT}). Furthermore, Ag and Mg are known to interact strongly \cite{Polmear_Role_1987}. The addition of Ag attracts Mg to the $\{111\}_\mathrm{Al}$ planes, which greatly promotes the $\Omega$ phase and suppresses the S phase. This is evident by the existence of a mono-layer of Ag associated with a mono-layer of Mg at the coherent interface of $\Omega$-Al; such interfacial phase can independently exist at the early stages of precipitation \cite{Kang2014}. Precipitates nucleate from solute clusters, and hence the determination of the location of the clusters is important. As Ag decomposes quickly from the solid solution and interacts strongly with other solute elements and quenched-in vacancies, the preference of Ag aggregation provides a special kind of heterogeneous nucleation site. The nucleation sites are strongly biased on $\{111\}_\mathrm{Al}$ planes, thus giving Ag the ability to modify subsequent precipitation. Interestingly, Ag also aggregates on \{0001\} planes in Mg (\{0001\}/\{111\} planes are the close-packed planes in HCP/FCC), stimulating precipitation in magnesium alloys \cite{zhang2016role,Zhu2016}. It is not the purpose of this paper to unify the microalloying mechanisms of Ag in aluminium, but we hope the present study on the Al-Ag binary system will provide a useful reference to the phase transformations of complicated aluminium alloys containing Ag. Particularly, the preference of specific crystallographic planes for Ag aggregation may shed light on its microalloying effects in aluminium.\par


\section{Conclusion}
We performed scanning transmission electron microscopy to examine the phases and phase transformations in an Al-1.68 at.\% Ag alloy. The energetics of Ag clustering within aluminium were studied by density functional theory. The main conclusions are as follows:

1. We discovered a new precipitate phase which we named $\zeta$ in the Al-Ag system. The $\zeta$ phase is an intermediate precipitate phase between GP zone $\epsilon$ and $\gamma^\prime$/$\gamma$ in the Al-Ag precipitation sequence. The structure of $\zeta$ is characterised by the long range ordering of bilayers enriched in Al and Ag on alternative $\{111\}_\mathrm{Al}$ planes. The $\zeta$ phase is coherent and displays alternating lattice displacements relative to the aluminium matrix in $\langle 111 \rangle_\mathrm{Al}$. The composition of $\zeta$ is close to AgAl.

2. Small Ag enriched clusters are formed during quenching and they grow into GP zones $\epsilon$ with inhomogeneous Ag distribution during ageing at 200\degree C. This chemical inhomogeneity is caused by the Ag aggregation on $\{111\}_\mathrm{Al}$ planes with favourable spacing. GP zone $\epsilon$ with a local ordering of Ag solute may transform into $\zeta$, particularly via {\em in situ} annealing of a TEM sample with induced vacancies.

3. The fast decomposition from the solid solution and the preferred $\{111\}_\mathrm{Al}$ planes for aggregation are intrinsic properties of Ag in aluminium. It provides heterogeneous nucleation sites on $\{111\}_\mathrm{Al}$ planes when Ag is microalloyed in aluminium alloys and fundamentally influences precipitation.


\section*{Acknowledgements}

The authors acknowledge funding from the Australian Research Council (LE0454166, LE110100223), the Victorian State Government and Monash University for instrumentation, and use of the facilities within the Monash Centre for Electron Microscopy. LB and NM acknowledge the financial support of the Australian Research Council (DP150100558). Authors also gratefully acknowledge the computational support from Monash Sun Grid cluster, the National Computing Infrastructure and Pawsey Supercomputing Centre. ZZ is thankful to Monash University for a Monash Graduate Scholarship, a Monash International Postgraduate Research Scholarship and a Monash Centre for Electron Microscopy Postgraduate Scholarship. ZZ is indebted to Matthew Weyland for his training on tilt series, Scott Findlay for his help on image simulations, Tianyu Liu for a Gatan Digital Micrograph script, Peter Miller for X-ray absorption analysis, Jonathan Peter for discussing geometric phase analysis, Xiang Gao for alloy casting, Ian Polmear for advice, and Yiqiang Chen and Xuan Chen for reading the manuscript.\par

\bibliographystyle{elsarticle-num}

\clearpage

\begin{table}
\def\arraystretch{0.6}
\centering
\begin{tabu} to 0.8\textwidth {X[c] X[c] X[c] X[c] X[c] X[c] X[c]}
\toprule
\multirow{2}{*}{Site} & \multicolumn{3}{c}{Experiment} & \multicolumn{3}{c}{DFT} \\ \cmidrule{2-4} \cmidrule{5-7}
 & x & y & z & x & y & z \\
 \midrule
Al(1)  & 1/3 & 2/3 & 0.086 & 1/3 & 2/3 & 0.081 \\
Al(2)  & 2/3 & 1/3 & 0.165 & 2/3 & 1/3 & 0.163 \\
Al(3)  & 2/3 & 1/3 & 0.417 & 2/3 & 1/3 & 0.414 \\
Al(4)  & 0 & 0 & 0.500 & 0 & 0 & 0.496 \\
Al(5)  & 0 & 0 & 0.741 & 0 & 0 & 0.747 \\
Al(6)  & 1/3 & 2/3 & 0.836 & 1/3 & 2/3 & 0.830 \\ \hline
Ag(1) & 0 & 0 & 0 & 0 & 0 & 0 \\
Ag(2)  & 0 & 0 & 0.248 & 0 & 0 & 0.244 \\
Ag(3)  & 1/3 & 2/3 & 0.336 & 1/3 & 2/3 & 0.333 \\
Ag(4)  & 1/3 & 2/3 & 0.580 & 1/3 & 2/3 & 0.577 \\
Ag(5)  & 2/3 & 1/3 & 0.668 & 2/3 & 1/3 & 0.667 \\
Ag(6)  & 2/3 & 1/3 & 0.917 & 2/3 & 1/3 & 0.910 \\
\bottomrule
\end{tabu}
\caption{Atomic coordinates of the precipitate $\zeta$ phase. The listed coordinates are fractional in respect to the simplest trigonal cell with space group of $P3$. It is equivalent to a trigonal cell with a space group of $R\bar{3}m$ (hexagonal axis) and Wyckoff positions of Ag at (0 0 0.878) and Al at (0 0 0.375). The experimental parameters are $a_\mathrm{exp}$=$2.88 \pm 0.05$ \AA{} and $c_\mathrm{exp}$=$27.35 \pm 0.05$ \AA. The DFT-optimised parameters are $a_\mathrm{DFT}$=2.97 \AA{} and $c_\mathrm{DFT}$=26.88 \AA{} for the bulk $\zeta$ phase and $a_\mathrm{DFT}^{emb}$=2.92 \AA{} and $c_\mathrm{DFT}^{emb}$=27.26 \AA{} for the embedded $\zeta$ precipitate phase. The uncertainty in the experimentally determined z coordinates is 0.005.}
\label{table:site}
\end{table}
\clearpage

\begin{table}
\centering
\begin{tabular}{l c c }
\toprule
Number of Ag Atoms & Configurations & $E_\mathrm{F}^\mathrm{Ag}$ (meV)\\
\midrule
1 Ag (Solid Solution) & N/A & 89 \\\hline

{2 Ag} & $1^\mathrm{st}$ nearest neighbour & 44 \\\cline{2-3}
& $2^\mathrm{nd}$ nearest neighbour & 99\\\hline

{3 Ag} & $\{001\}_\mathrm{Al}$ & 27 \\\cline{2-3}
& $\{110\}_\mathrm{Al}$ & -1 \\\cline{2-3}
& $\{111\}_\mathrm{Al}$ & -1 \\\hline

{Ag plane} & $\{001\}_\mathrm{Al}$ & 431 \\\cline{2-3}
& $\{110\}_\mathrm{Al}$ & 66 \\\cline{2-3}
& $\{111\}_\mathrm{Al}$ & -65 \\
\bottomrule
\hline
\end{tabular}
\caption{DFT calculations for the preference of Ag clustering in Al matrix. $E_\mathrm{F}^\mathrm{Ag}$ is the formation energy per Ag atom. $1^\mathrm{st}$ nearest neighbour stands for two Ag atoms in the nearest neighbour configuration in a $\langle 110 \rangle_\mathrm{Al}$ direction in FCC Al lattice. Similarly, the $2^\mathrm{nd}$ nearest neighbour is two Ag atoms next to each other in a $\langle 001 \rangle_\mathrm{Al}$ direction.}
\label{table:DFT}
\end{table}
\clearpage

\begin{figure}
\centering
 \includegraphics[width=0.8\textwidth]{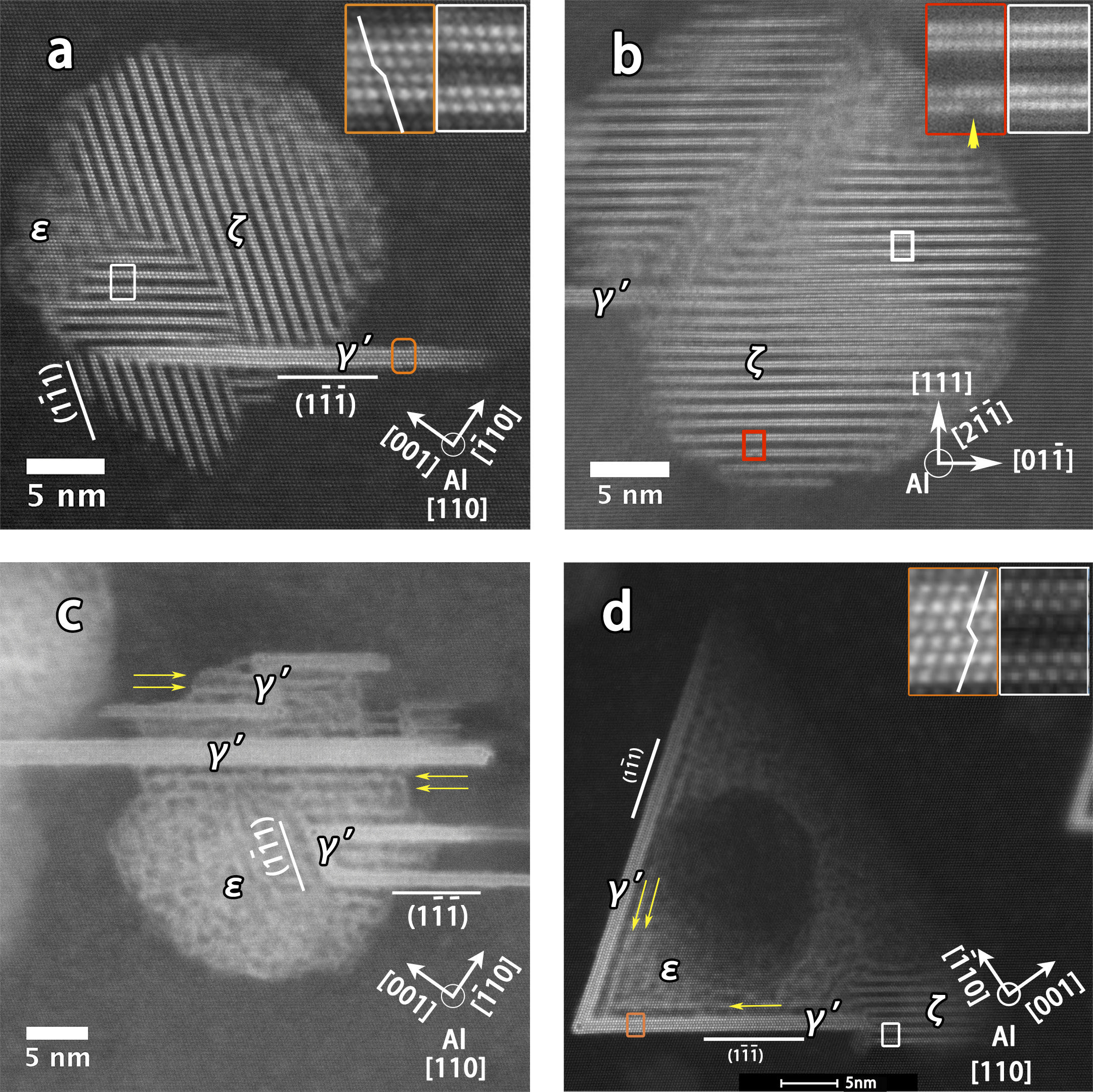}
 \caption{HAADF-STEM images of the transformed GP zones. (a) A bi-layered phase formed on $\{111\}_\mathrm{Al}$ planes and viewed along $\langle 110 \rangle_\mathrm{Al}$, where the white framed region shows that the bi-layered phase has an ABCABC stacking and the orange framed region shows the characteristic stacking fault of $\gamma^\prime$; (b) a bi-layered phase viewed along a $\langle 112 \rangle_\mathrm{Al}$ direction, where the white framed region shows the uniformly enriched Ag layers while the red framed region shows non-uniformly enriched Ag layers as indicated by a yellow arrow; (c) $\gamma^\prime$ plates formed inside a GP zone introducing ordering as indicated by yellow arrows; (d) a bi-layered phase formed at the tail of a $\gamma^\prime$ assembly, as viewed along $\langle 110 \rangle_\mathrm{Al}$. Images (a-c) were from the sample that underwent {\em in situ} annealing as shown in Fig.~\ref{fig:insitu}. Image (d) was from a sample that underwent a conventional heat treatment: water quenched and aged at 200\degree C for 2 h without {\em in situ} annealing.}
 \label{fig:Z}
\end{figure}


\end{document}